# External Dynamic InTerference Estimation and Removal (EDITER) for low field MRI

**Sai Abitha Srinivas[1,2*], Stephen F Cauley[3,4], Jason P Stockmann[3,4], Charlotte R Sappo[1,2], Christopher E Vaughn[1,2], Lawrence L Wald[3,4,5], William A Grissom[1,2,6,7] and Clarissa Z Cooley[3,4*]**

[1]Vanderbilt University Institute of imaging science, Nashville, TN, United States, [2]Department of Biomedical Engineering, Vanderbilt University, Nashville, TN, United States, [3]Harvard Medical School, Boston, MA, United States, [4]Dept. of Radiology, Massachusetts General Hospital, Athinoula A Martinos Center for Biomedical Imaging, Boston, MA, United States, [5]Harvard-MIT Division of Health Sciences and Technology, Cambridge, MA, United States, [6]Department of Electrical Engineering, Vanderbilt University, Nashville, TN, United States, [7]Department of Radiology, Vanderbilt University, Nashville, TN, United States

*To whom correspondence should be addressed:
sai.abitha.srinivas@vanderbilt.edu, clarissa@nmr.mgh.harvard.edu

**Abstract**

**Purpose:** Point-of-care MRI requires operation outside of a Faraday shielded room normally used to block image-degrading electromagnetic Interference (EMI). To address this, we introduce the EDITER method, an external sensor-based method to retrospectively remove image artifacts from time-varying external interference sources.

**Theory and Methods:** The method acquires data from multiple EMI detectors (tuned receive coils as well as electrodes placed on the body) simultaneous with the primary MR coil during and between image data acquisition. We dynamically calculate impulse response functions that map the data from the detectors to the artifacts in the kspace data, then remove the transformed detected EMI from the MR data. Performance of the EDITER algorithm was assessed in phantom and *in vivo* imaging experiments in an 80mT portable brain MRI in a controlled EMI environment and with an open 47.5mT MRI scanner in an uncontrolled EMI setting.

**Results:** In the controlled setting, the effectiveness of the EDITER technique was demonstrated for specific types of introduced EMI sources with up to a 97% reduction of structured EMI and up to 76% reduction of broadband EMI. In the uncontrolled EMI experiments, we demonstrate EMI reductions of 37% with a single pickup coil and 89% with a single electrode and up to 99% with both.

**Conclusion:** The EDITER technique is a flexible and robust method to improve image quality in portable MRI systems with minimal passive shielding. This could reduce the reliance of MRI on shielded rooms and allow for truly portable MRI with specialized compact POC scanners.

## Introduction

There is growing interest in low-field, low-cost, and portable MRI scanners for democratizing MRI and extending its use to unconventional locations [1-3]. Specialized brain scanners are a particularly attractive target for this role since brain MRI outshines other modalities but can be underutilized due to cost or lack of access [4]. Simplified low-field portable brain scanners could be used as point-of-care (POC) [5 - 18] instruments in hospital EDs and ICUs or as easy-to-site scanners in underserved and unconventional locations (hospitals in the developing world, rural clinics, urgent care clinics, sports arenas, etc.).

However, the surrounding RF shielded room ("Faraday shield") is an essential element of conventional MRI scanners which shields the MR receive coils from external sources of Electro Magnetic Interference (EMI). Nuisance electromagnetic signals are generated from a wide variety of external sources (including nearby electronic equipment) and radiation at frequencies close to the Larmor frequency can contaminate the MR signal through electromagnetic induction in the primary MRI imaging coils. If unmitigated, the interference can significantly degrade the image quality and reduce diagnostic utility. Although RF shielded rooms are successful at passively attenuating EMI, their necessity renders the system non-portable and can preclude flexible POC use. To enable truly portable MRI devices, alternative approaches to EMI suppression are needed.

One approach is to apply local passive conductive shielding in and around the portable scanner (as opposed to the room). Of course, for a brain MRI, these RF shields must have an opening to accommodate the subject's body, but may be built around the outside of the magnet [6] or directly exterior to the RF coil [7]. For these low-field systems (60mT and 80mT), the shields' opening is small compared to the Larmor frequency wavelength, and therefore effectively attenuates interference for imaging phantoms contained within the shield. However, when imaging a human subject, EMI is conducted by the body parts external to the magnet and shield and is picked-up by the RF coil. This requires that passive shielding encompass the entire body of the patient, for example, using conductive mesh cloth draped over the patient [6]. For maximum effectiveness, the conductive cloth must be completely sealed around the body, which can be uncomfortable and limit access to the patient.

Post-processing approaches to EMI mitigation are more flexible and are less intrusive to the subject. An early method by Letcher [19] reduced EMI artifacts with Weiner filter deconvolution. The filter transfer function was formed from the power spectrum of the MR data and "noise" data (calibration data) which were sampled by the primary MR receiver during dummy cycles in the acquisition. Similar methods using additional external coils as EMI detectors have also been proposed [20–22] and are likely in use in at least one commercial device [5] but there is little literature demonstrating the approaches' efficacy. In these methods, calibration data is sampled to form complex transfer functions correlating the external EMI coil data to the primary MRI coil data. The reliance on calibration data from pre-scan data alone could hinder the ability to mitigate time-varying EMI sources and could increase scan time if the pre-scans needed regular updates. We note that EMI mitigation is also needed for other biomedical measurements, such as ECG and EEG, where adaptive filter algorithms have been demonstrated with electrode data [21,22].

We propose an approach to dynamic EMI correction using data that is acquired simultaneously during the scan from EMI detectors external to the imaging volume and the primary MR coil. The EMI detectors include external RF coils as well as electrodes attached to the subject. We term our method: External Dynamic InTerference Estimation and Removal (EDITER). Our approach is based on a dynamic model comprising a kernel that relates the external detectors' signal to the artifact's appearance in the recording of primary imaging coil. The estimated EMI contamination is removed from the primary MR data in postprocessing. Previously, we demonstrated the use of dynamic calibration data collected at the end of each acquisition TR period (acquired during deadtime) to calculate a new transfer function for each line of k-space [25]. Here, we show that this calibration data is not needed and that the model can be fit directly using data acquired during NMR signal reception. Since the signal is simultaneously acquired by both the EMI detectors and the primary MR coil and there are no additional acquisition periods beyond what is needed for imaging, the method can be added to any sequence with no modifications. Similar to TGRAPPA [26, 27] we group detector acquisition lines with similar signal properties to build a time-varying model, which is more robust to noise compared to using single acquisition lines [28]. This method provides larger convolution windows to correlate the EMI data which yields more accurate impulse response functions and better EMI removal. We test the method's ability to improve image quality in scenarios where EMI is not stationary thus potentially allowing operation of the portable scanners in EMI unfriendly environments without the traditional Faraday shielded room or Faraday shield bedding.

EDITER was validated in two low-field MRI systems at different centers. Phantom imaging experiments were performed with an 80mT portable head scanner [7] in a shielded room with controlled EMI sources introduced. The second low field system was an open design head scanner with planar gradients at 47.5mT. We applied the proposed algorithm in-vivo without a shielded room to test its effectiveness of the method in uncontrolled EMI settings.

## Theory

With the EDITER method, we propose a generalized model that can dynamically adjust to time-varying external noise sources. The model allows for simultaneously acquired data from multiple EMI detectors to be regressed from the primary MR coil data. Here, a linear relationship is assumed between the kspace signal measured by the primary receive coil $s(k_x, k_y)$, the unwanted EMI on the imaging coil $e'(k_x, k_y)$, and the desired EMI-free kspace data $s'(k_x, k_y)$:

$$s(k_x, k_y) = e'^{(k_x,k_y)} + s'^{(k_x,k_y)} \qquad (eq. 1)$$

To allow for accurate estimation and removal of the EMI, we assume data is available from $N_c$ external detectors $e_i(k_x, k_y), i = 1, \ldots, N_c$. A linear convolution model along the readout ($k_x$) and phase encoding ($k_y$) directions relates the EMI observed by the primary imaging coil to that observed by the external detectors:

$$e'(k_x, k_y) = \sum_{i=1}^{N_c} e_i(k_x, k_y) * h_i(k_x, k_y) \qquad (eq.2)$$

Each impulse response function is assumed to have limited spectral support, i.e., $h_i(k_x, k_y) = 0, |k_x| > \Delta k_x \; or \; |k_y| > \Delta k_y$. In the most restrictive case, $\Delta k_x = 1, \Delta k_y = 1$, (eq. 2) represents a scalar combination of the detector coils:

$$e'(k_x, k_y) = \sum_{i=1}^{N_c} e_i(k_x, k_y) \cdot h_i \qquad (eq.3)$$

The linear convolutional form described in (eq. 2) can be incorporated into (eq. 1) and written in matrix form as: $\vec{s} = \vec{e}' + \vec{s}' = E\vec{h} + \vec{s}'$. The impulse response vector $\vec{h} \in \mathbb{C}^{N_c \Delta k_x \Delta k_y \times 1}$ is a concatenation of the spectral components for the impulse response functions and the EMI convolution matrix $E \in \mathbb{C}^{N_F \times N_c \Delta k_x \Delta k_y}$ is a block-Toeplitz matrix mapping the EMI detector coil data to $N_F$ observations from the primary receive coil. An illustration of fitting impulse response functions is shown in Figure 1A. In the case of temporally static EMI, a single set of impulse response functions would be valid across the full extent of kspace and all available data could be used during the fit, i.e. $N_F = N_{k_x} \cdot N_{k_y}$. While $k_x$ and $k_y$ coordinates are used in the present mathematical treatment for familiarity and convenience to index data points within the readouts ($k_x$) and between readouts ($k_y$), we note that the method does not depend in any way on the specific k-space sampling trajectory used.

The least squares solution $\vec{h} = E^{\dagger}\vec{s}$ is used to fit the model and the EMI mitigated data can be produced as: $\vec{s}' \approx \vec{s} - E\vec{h}$. It is important to note that this method assumes low correlation between the spectral content of the image and noise sources. In the unlikely event that such correlation does exist, this would merely lead to smoothly varying sensitivity loss across the image due to the limited support assumed for each $h_i(k_x, k_y)$.

We generalize the method to scenarios where EMI sources are time-varying by assuming limited temporal windows (e.g. one or more successively acquired phase encode lines) and fit different impulse response functions for each temporal instance. As the number of observations in each temporal window becomes small, the estimation robustness will be degraded. To minimize this effect, we dynamically bin the data into larger temporal windows that have consistent EMI patterns. This is accomplished by first constructing a matrix $H$ that contains the $\vec{h}$ vectors generated for the different temporal windows $N_w$. $\Delta k_y$ used to construct the H matrix is restricted by $N_w$ used. Here, we use $N_w$=1 therefore, $\Delta k_y$=1. We then construct a matrix $C$ by autocorrelating the normalized matrix $H$. Consistent EMI sources will lead to similar impulse response functions which in turn will produce high levels of correlation. The binning locations can be determined with standard clustering approaches [30] such as thresholding the correlation matrix $C$ to form $C_{\text{threshold}}$ as shown in Figure 1B. A final pass of the method is then performed to estimate and remove EMI from each dynamically determined temporal window. Here, there is no restriction on $\Delta k_y$. Pseudocode for the generalized algorithm is provided below:

1. **Distribute PE lines into $N_W$ small temporal windows of size: $W = N_{PE}/N_W$.**
2. **For each temporal window $1 \leq k \leq N_W$:**
    a. **Formulate convolution matrix: $E_{(k)} \in \mathbb{C}^{(N_{k_x}W)\times N_c \Delta k_x \Delta k_y}$, see Figure 1A.**
    b. **Arrange primary receive signal into vector: $\vec{s}_{(k)} \in \mathbb{C}^{(N_{RO}W)\times 1}$.**
    c. **Compute impulse response functions: $\vec{h}_{(k)} = E^{\dagger}{}_{(k)}\vec{s}_{(k)}$.**
3. **Calculate impulse response correlation matrix: $C_{i,j} = \langle \vec{h}_{(i)}, \vec{h}_{(j)} \rangle, 1 \leq i,j \leq N_W$.**
4. **Cluster temporal windows into $N_G$ groups, e.g. using [20], see Fig. 1B.**
5. **For each temporal cluster $1 \leq g \leq N_G$:**
    a. **Formulate convolution matrix $E_{(g)}$.**
    b. **Arrange primary receive signal into the vector: $\vec{s}_{(g)}$.**
    c. **Compute impulse response functions: $\vec{h}_{(g)} = E^{\dagger}{}_{(g)}\vec{s}_{(g)}$.**
    d. **Determine EMI mitigated signal: $\vec{s}'_{(g)} = \vec{s}_{(g)} - E_{(g)}\vec{\mathrm{h}}_{(g)}$.**

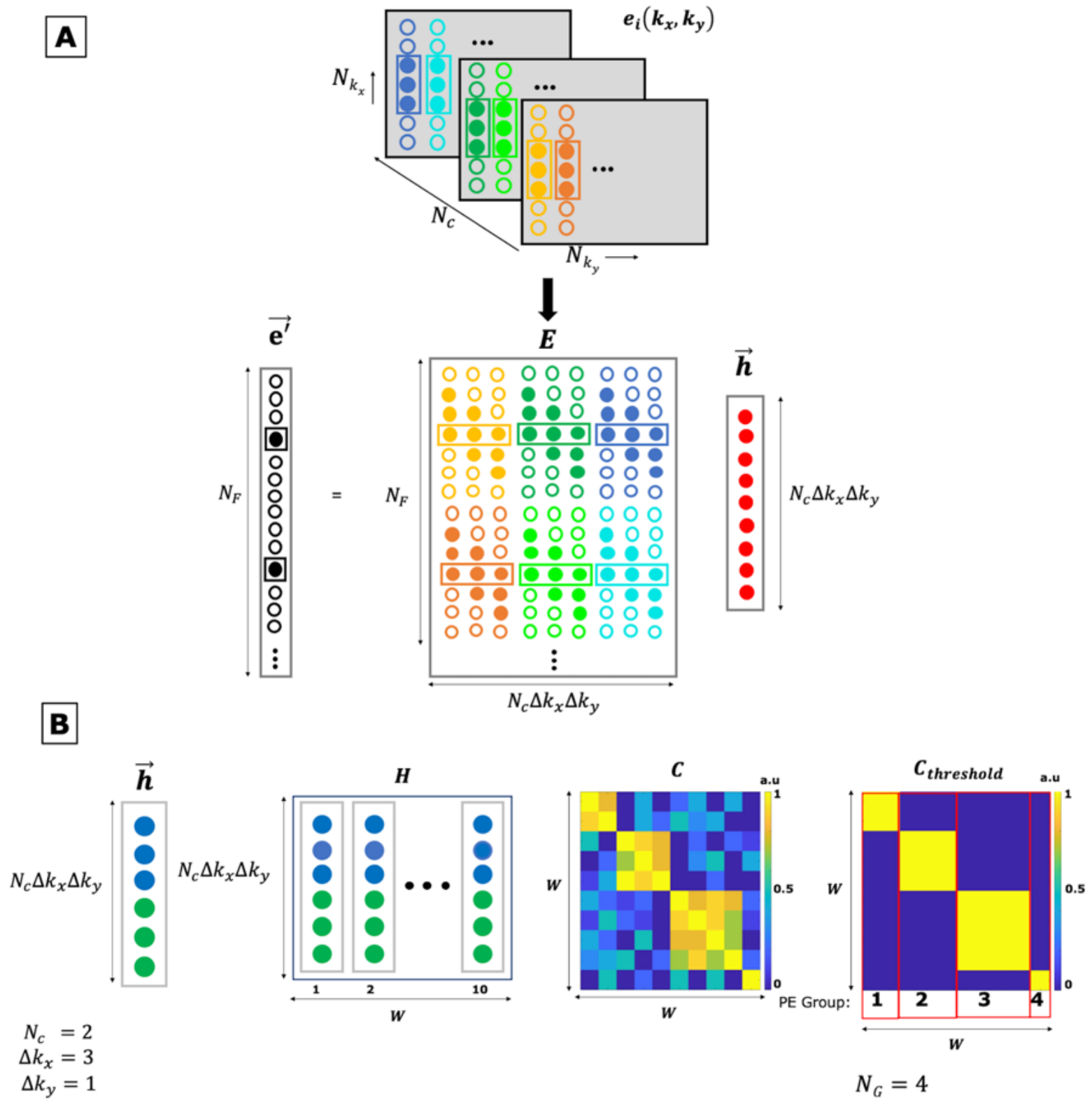

***Figure 1: EDITER EMI estimation overview.*** *A) Fitting Impulse Responses. Top: The k-space data of the external EMI detectors* $e(k_x, k_y)$ *is acquired simultaneously with the imaging data from all the EMI detectors (*$N_c$*). Bottom: The EMI present in the primary image,* $\vec{e'}$*, is estimated with the impulse response function* $\vec{h}$ *and the EMI convolution matrix* $E$*.* $E$ *comprises of shifted copies of the phase encode lines with a convolution window size* $\Delta k_x$ *= 3 and* $\Delta k_y$ *= 1 for all EMI detectors. The impulse response is obtained using the inverse of the EMI convolution matrix* $E$ *and the primary MR coil data. B) Dynamic Phase Encode Binning method. Left to right – First, we form an impulse response function* $h$ *for each temporal group of PE lines individually (W). Here shown for two EMI detectors (*$N_c$ *= 2, blue and green) and a convolution window* $\Delta k_x$ *= 3 and* $\Delta k_y$ *= 1. The individual impulse response functions are concatenated to form the* $H$ *matrix with size (Nc ×* $\Delta k_x$ *×* $\Delta k_y$*) × W (Here, W = 10). We then form a correlation matrix* $C = H'H$ *wherein* $H$ *is normalized. This matrix is then clustered using a thresholding value of r = 0.5 to form a binary matrix* $C_{threshold}$ *that is used to group phase encode lines. Here, we arrive at four groups (*$N_G$ *= 4) with two, three, four and one phase encode lines each.*

## Methods

**Experiments in controlled EMI settings in an 80mT portable low-field brain MRI**: A portable, head-only low-field MRI scanner at MGH [7] is used to demonstrate our method in phantoms. The portable system is based on a compact 122 kg, 80 mT permanent magnet Halbach cylinder that requires no power or cooling. The permanent magnet produces a built-in field gradient for readout encoding – eliminating the readout gradient system and thus reducing the system's power and cooling needs. The built-in readout gradient necessitates the use of spin-echo sequences and frequency-swept RF pulses to cover the full Larmor frequency bandwidth [31, 32]. Recently published *in vivo* imaging results with this scanner were acquired in a stationary RF shielded room to attenuate EMI [32].

To validate EDITER, we acquired phantom data in the shielded room with introduced EMI sources to fully control the EMI contamination. This gave us the ability to acquire "ground truth" EMI-free images simply by turning off the sources and also generate time-varying sources. Four imaging experiments were performed with the following sources introduced: 1) A single coherent frequency source was generated with a wire loop connected to a frequency generator (FG) producing a 10 Vp-p sine wave at the Larmor frequency (3.38 MHz) of the magnet (positioned at the end of the patient table). 2) A stepper motor (SM) was used to produce robust sporadic EMI at multiple frequency bands (positioned behind the magnet). 3) Three cascaded 1W RF power amplifiers (RFPA) (Mini-circuits ZHL- 3A+, NY, USA) connected to a untuned loop were used as the broadband (BB) EMI source (positioned underneath the scanner). 4) Time-varying EMI was introduced by manually switching SM and BB EMI sources during the scan.

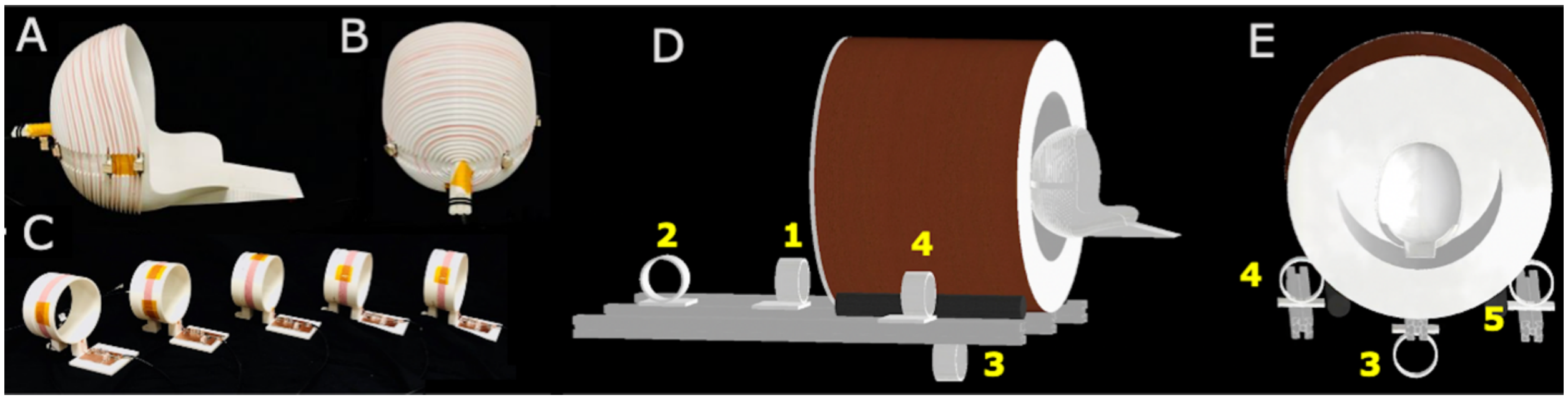

***Figure 2: Experimental setup for portable 80mT scanner. (A-B)*** *The single channel primary MR spiral volume Tx/Rx coil tuned to 3.38MHz for the portable 80 mT scanner with 25 KHz BW. C) Five external EMI detector coils were placed outside the magnet tuned to the Larmor frequency 3.38MHz with a 25KHz BW to mimic the frequency response of the primary MR coil. D-E) Placement of the EMI detector coils around the outside of the portable 80mT head scanner. The yellow numbers indicate the detector channel number.*

Five identical EMI detectors were built from 10-turn coils wound on 3D printed formers (OD = 8 cm) as shown in figure 2C. These EMI detectors were tuned and matched to the Larmor frequency of the scanner (3.38 MHz) with a bandwidth of 25 KHz to achieve a similar frequency response to the primary RF coil. Two detectors were used with 50 ohm 37 dB gain pre-amplifiers (MITEQ P/N AU 1583, NY, USA) and three detectors were used with 50 ohm 24 dB gain preamplifiers (Mini-circuits ZHL-500LN+, NY, USA). EMI detector 1 and EMI detector 2 were placed behind the sides of the scanner. EMI detector 3 was placed underneath the scanner. EMI detector 4 and 5 were placed by the sides of the scanner. EMI detectors 1,3, 4 and 5 were oriented along the B1 direction. EMI detector 2 was oriented orthogonal to B1 to acquire EMI characteristics in an additional direction. The placement of these detectors is illustrated in figure 2D and 2E.

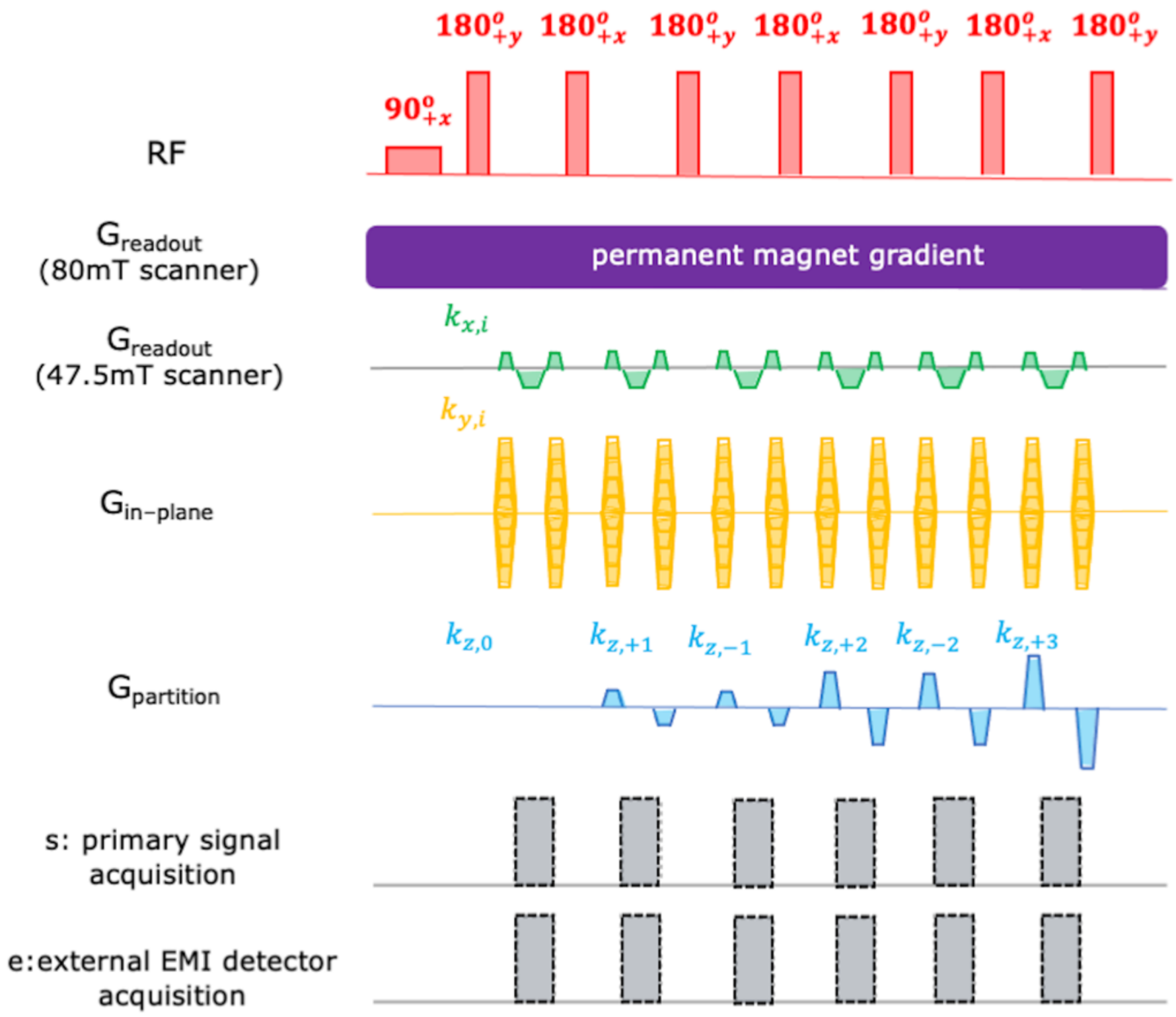

***Figure 3:*** *The 3D RARE (Rapid Imaging with Refocused Echoes) pulse sequence used for PD-weighted imaging in both the scanners. The $G_{readout}$ gradient was the built-in permanent magnet encoding field for the portable 80mT low field head scanner and therefore was continuously applied throughout the acquisition. A conventional pulsed readout gradient as shown in the fourth row was played out in the 47.5mT open low field system. The $G_{in\text{-}phase}$ and $G_{partition}$ phase encoding blips were applied using dedicated gradient coils for both systems. The signal acquisition alternated between the narrow "FID echoes" and wider "spectral echoes" in the portable 80mT scanner. Only "FID echoes" were generated in the 47.5mT low field scanner using hard pulse excitation and refocusing. Data was acquired simultaneously in the primary MR coils and the external EMI coils.*

Phantom experiments were performed using an MR Solutions (Guilford, UK) console with 32 Rx channels and 4 Tx channels. Other hardware included 2 AE Techron 7224 gradient amplifiers (Elkhart, IN, USA) and a 2kW peak-power RF power amplifier (Tomco Technologies model BT02000-AlphaS-3MHz, Stepney, Australia). Data was acquired using a single channel RF transmit/receive coil for primary MR data shown in Figure 2A and 2B [33,34] with a 50 ohm 37 db gain pre-amplifier (MITEQ P/N AU 1583, NY, USA) and 24 dB second stage amplifier (Mini-circuits ZHL-500LN+, NY, USA). Experiments were performed with a 1 cm thick 3D printed brain slice phantom.

A 2D multi-echo Rapid Acquisition and Relaxation Enhancement (RARE) volumetric spin echo sequence [35] as shown in Figure 3 was used to obtain the data shown in Figure 6. The sequence parameters used were - Resolution = 2.2mm x1.3mm, Matrix size = 512 x 101.

EDITER correction was performed on each EMI source data set. The correction was performed with each detector coil individually and with the combined detection data. To determine the impulse response vectors, $\vec{h}$, we used an EMI convolution window size of $\Delta k_x = 7$ and $\Delta k_y = 1$. In these controlled experiments, we compare the corrected and uncorrected images and quantify the EMI mitigation performance with 2 different metrics. Metric 1 is an RMSE comparison of the image space residual compared to ground truth. Metric 2 estimates the % EMI removal using the standard deviation in a region outside the object in the corrected and uncorrected images (equation 4).

$$EMI_{\substack{residual\ wrt \\ uncorrected}} = \left|\frac{\sigma_{UN} - \sigma_C}{\sigma_{UN}}\right| \times 100\ \% \qquad (eq. 4)$$

In equation 4, $\sigma_{UN}$ and $\sigma_C$ represent the standard deviation of an EMI region in the uncorrected image and the corrected images respectively. The EMI region was such that no part of the object was included in the calculation.

***Experiments in uncontrolled settings*** **in a 47.5 mT open low-field system:** *In vivo* human brain imaging was performed at Vanderbilt University with the 47.5 mT permanent magnet MRI scanner shown in Figure 4 (Sigwa MRI, Boston, MA, USA) [36]. The Sigwa scanner weighing 13380 kg is an open, biplanar permanent magnet imaging system with an 80 cm gap and 2 m x 1.4 m footprint. The $B_0$ field has 18.7 ppm homogeneity over the center 40 cm DSV and 5 Gauss line approximately 2.5 m from isocenter. This high homogeneity enabled the use of hard pulse excitations instead of the frequency-swept pulses used for the portable 80 mT scanner described above. Mounted to the magnet pole pieces are 3-axis planar gradient coils with < 6% linearity in a 40 DSV and < 1.25% linearity in a 20 cm DSV. The scanner is controlled with a Tecmag Redstone console (Houston, TX, USA) with two transmit and three receive channels. The gradients are driven by three AE Techron 2120 amplifiers (Elkhart, IN, USA) and a 500W peak-power RF power amplifier is used (Tomco Technologies model BT00500-AlphaS, Stepney, Australia). *In vivo* images were acquired using a single channel Tx/Rx RF coil [8,23,24] similar to that used the portable head scanner (Figure 4A) and a 50-ohm 70 dB gain pre-amplifier (MITEQ P/N 1583 10057, NY, USA). The Sigwa scanner is not sited in an RF shield. The EMI contamination in these experiments came from uncontrolled environmental sources at Vanderbilt University Medical Center.

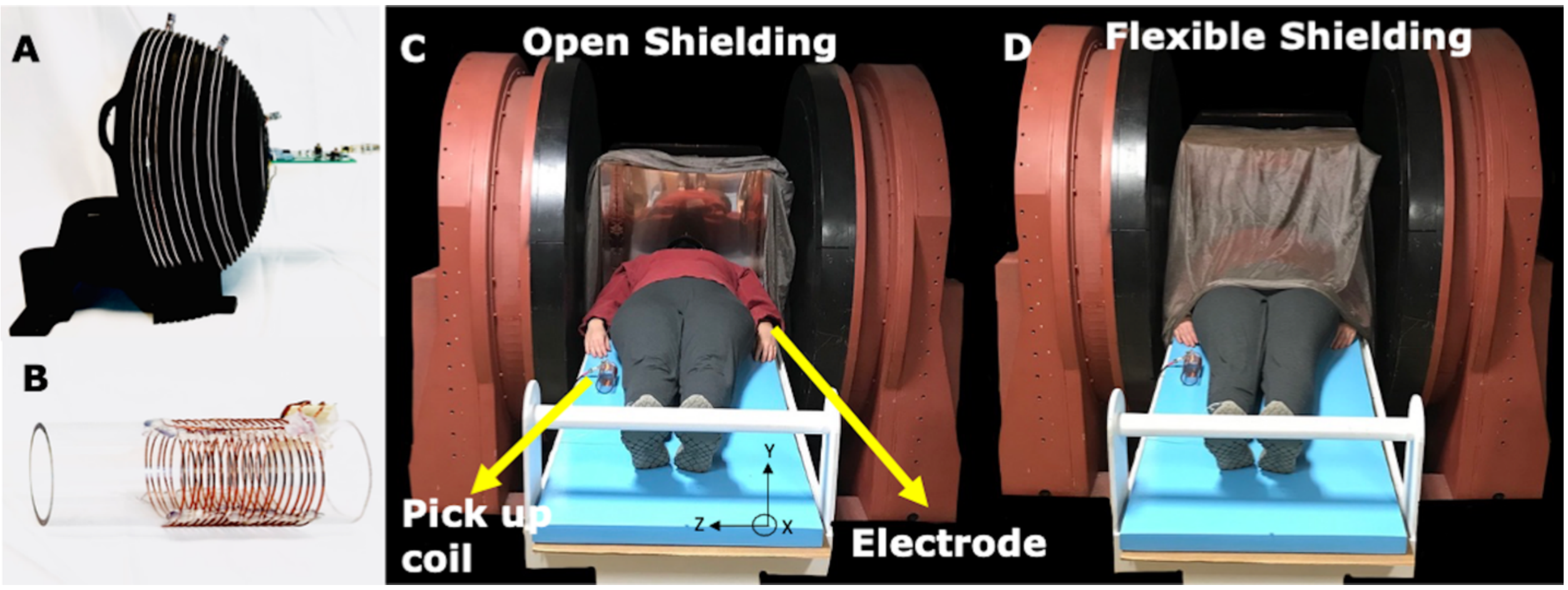

***Figure 4: Experimental setup for open 47.5mT scanner.*** *A) The primary MR coil used for imaging (same spiral design as Fig. 2) tuned to 2.075 MHz with 20 KHz BW. B) EMI detector #1 is a tuned 10 turn RF pick up coil with diameter = 5cm. EMI detector #2 is an ECG electrode (not shown). C-D) The open 47.5mT low field scanner shown with subject and the two EMI detectors, including the electrode attached to subject's wrist. Experiments were performed in 2 shielding configurations. C) The open shielding configuration included a copper box surrounding the head open on one side. D) The flexible shielding configuration included the addition of copper mesh, grounded to the copper box and draped over the subject.*

Similar to ECG and EEG noise cancellation methods [23, 24], we propose the use of electrodes as EMI detectors to directly measure the EMI that is coupled through the patient. In these *in vivo* experiments, a single electrode (EverOne ECG Monitoring) attached to the patient's wrist to measure EMI and serves as 1 of 2 of the EMI detectors. The second EMI detector is a pick-up coil with 10 evenly spaced turns on an acrylic former (OD = 5 cm) (Figure 4B). The coil is tuned and matched to the Larmor frequency of the scanner (2.07 MHz) with a similar bandwidth as the primary MR coil (20 KHz). The pick-up coil was positioned by the side of the patient oriented in the primary coil's B1 direction as shown in figure 4C and 4D. Both EMI detectors were connected to 50-ohm 70 dB gain pre-amplifiers (MITEQ P/N 1647, NY, USA).

*In vivo* scans were performed with 2 different passive shielding configurations within the magnet (Figure 4c-d). The "open" configuration included a head-sized copper box open on one side surrounding the primary helmet coil (Figure 4C). The "flexible shielding" configuration included the addition of conductive cloth /Faraday shield bedding (Less EMF, NY, USA) draped over the subject (Figure 4D).

The 3D RARE sequence of Fig. 3 was used to obtain *in vivo* scans in three healthy subjects (S1: male, 24 y/o, S2: female, 27 y/o, S3: male, 30y/o) after providing consent in accordance with the Vanderbilt university Institutional Review Board (IRB) guidelines. The sequence parameters were: Resolution = 1mm x 2mm x 9mm, Matrix size =128 x 97 x 23. For EMI mitigation, we first divided the data into 2D partitions with the FFT, and then applied the EDITER algorithm to the 2D datasets individually. We used an EMI convolution window size of $\Delta k_x$ = 7 and $\Delta k_y$ = 1 for correction in all subjects in both the open and shielded configurations. The percent EMI correction for *in vivo* images were evaluated using equation 4 (metric 2). Metric 1 is not calculable, as a ground truth is unavailable in these uncontrolled EMI experiments. Additionally, Table 1 details the standard deviation for corrected and uncorrected data for both the flexible and open shielding configurations for the three subjects.

The choice of convolution window widths, $\Delta k_x$, is analyzed by comparing the EMI correction performance as well as the computation time. The computation was done in MATLAB and consisted of loading and reshaping the data and running the EDITER algorithm for all the partitions in the 3D volume. We used MATLAB version 2020a on a MacBook Pro with a 16 GB ram running a 2 GHz Quad-Core Intel Core i5 processor for the computation.

## Results

### ***Controlled EMI settings***

The four datasets were acquired with the previously described EMI sources introduced: 1) FG, 2) SM, 3) BB, 4) time-varying SM + BB. Supporting information S1 shows the spectra of each EMI source measured by each coil in the experiment. Figure 5 shows the EMI correction correlation matrix (C) for each experiment. Figures 5A-C show the C matrices for the stationary EMI experiments (FG, SM, BB) showing no significant variation in the correlation between PE lines. Figure 5D shows the correlation matrix for the time-varying EMI source, where the stepper motor (SM) was manually turned "on" for the first half of the image and the broadband (BB) EMI source was turned "on" for the second half. Therefore, we see three separate groups of correlated PE lines ($N_G$ = 3) in Figure 5D (the SM period, $\tau_{SM}$, the BB period, $\tau_{BB}$, and the switching period, $\tau_{Gap}$).

Figure 6 shows the 2D phantom imaging results from the controlled source experiments, including a ground truth image which was acquired with all EMI sources turned off. The uncorrected EMI contaminated images are shown for the 4 acquisitions with the 4 introduced EMI sources. Difference images and RMSE between the uncorrected and ground truth data are also shown. We note that the EMI stripes in the images follow the magnet's built-in non-linear gradient iso-contours and thus do not appear linear using a general reconstruction method detailed in [7]. The EMI corrected images are shown for each source using all 5 EMI pickup coils. Supporting Information S2 shows the corrected images using each EMI detector individually to demonstrate the contribution of each detector. Figure 6 includes the residual of the corrected images and RMSE (metric 1) compared to the ground truth. Comparing the uncorrected and corrected image RMSEs, shows a reduction by the EMI mitigation method of 89.2%, 95.7%, 74.7%, and 93.3% in the 4 acquisitions. Comparing the uncorrected to corrected images directly shows a decreased EMI percentage of 96.6%, 97.3%, 76.2%, and 86.8% using metric 2 in equation 4.

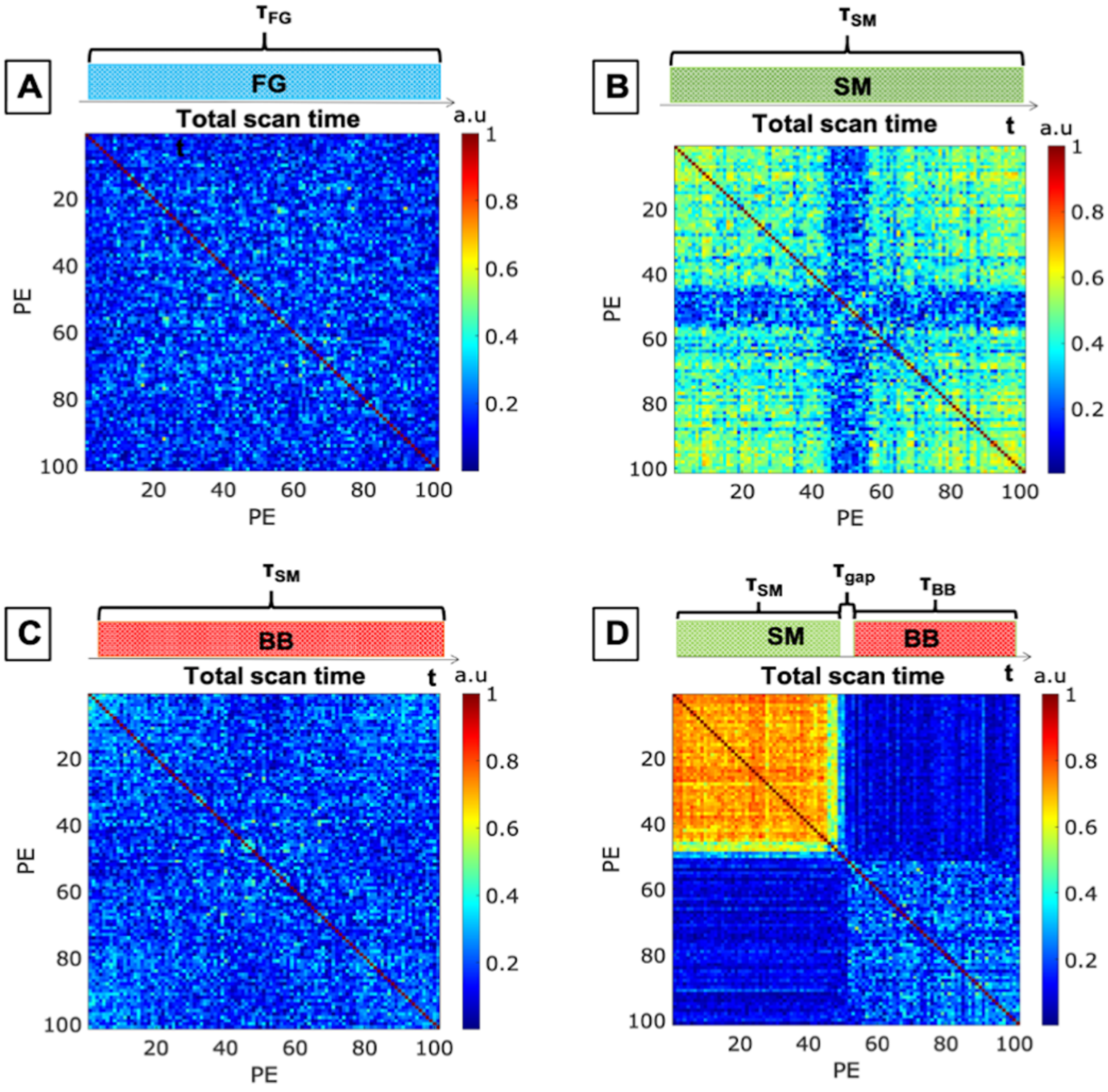

***Figure 5:*** *Correlation matrices C for the phantom experiments conducted on the portable 80mT scanner with the following introduced EMI. A) single band EMI produced by a function generator (FG). B) Narrowband EMI produced by a stepper motor (SM). C) Broadband EMI (BB). D) Dynamic switching between the SM and BB sources (narrowband stepper motor source on during the first half of the scan and broadband EMI source on during the second half).*

Although we applied the dynamic EMI correction algorithm, the stationary EMI source experiments (FG, SM, and BB) used single temporal window ($N_G$ = 1) for the correction because of the strong correlation of the impulse response across all PE lines (Figure 5). On the other hand, the time-varying EMI source (EMI 4: SM+BB) was corrected with $N_G$ = 3 temporal windows. To illustrate the effect of the dynamic vs. static EMI correction, we also show the SM+BB correction with a static impulse response in Figure 6. The static correction resulted in an NRMSE improvement of 93.3% and a percent EMI reduction of 86.8%, compared to 73.7% and 70.4% with the dynamic correction.

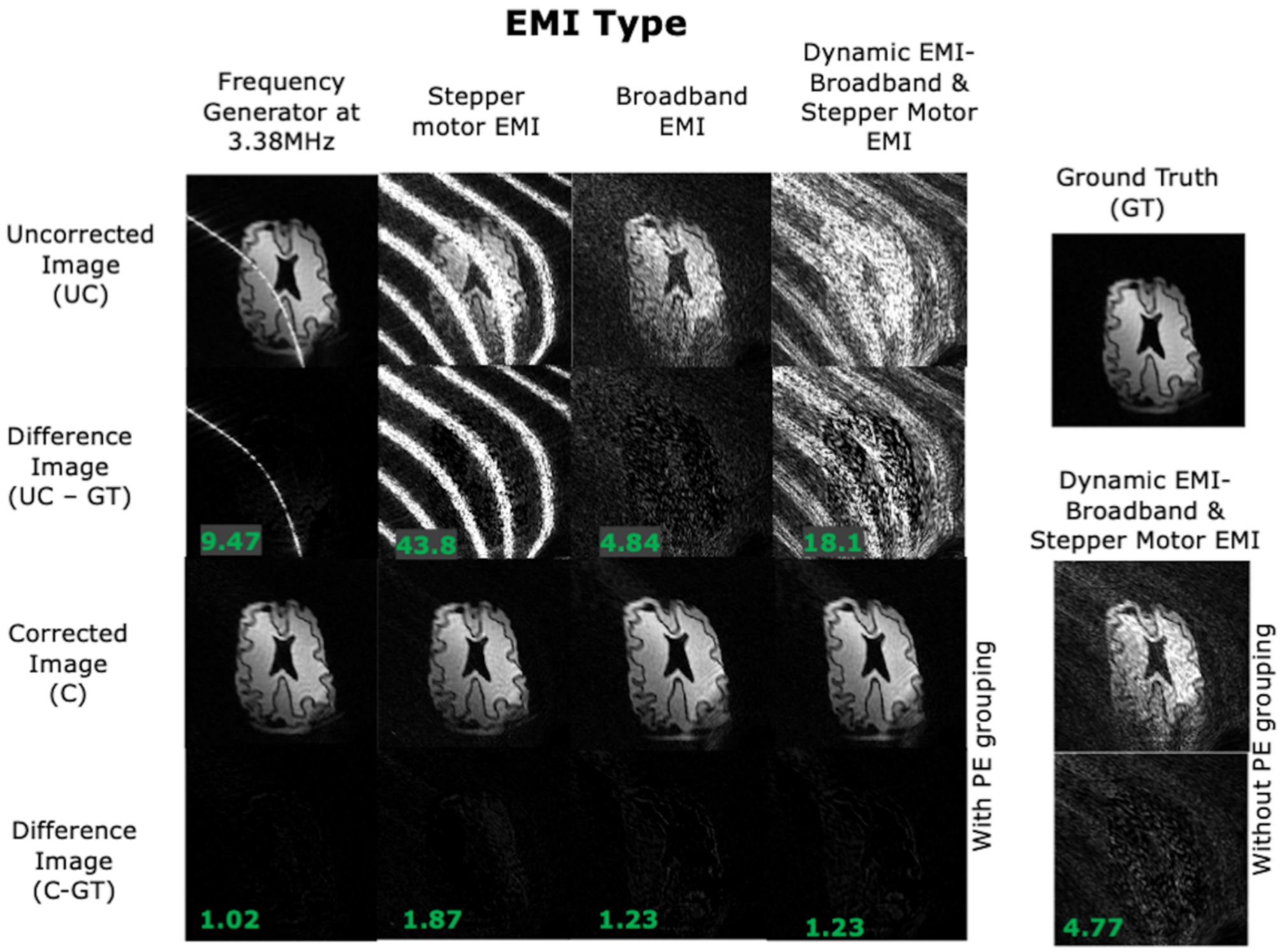

***Figure 6: EDITER-corrected 2D images of a brain slice phantom**. Results are shown for the 4 introduced EMI sources describe in Fig. 5. Top to bottom: Uncorrected image, Difference images between uncorrected image and ground truth, EDITER-corrected image using all EMI detectors and Difference images between corrected image and ground truth. The NRMSE (metric 1) of the difference images is shown in the bottom left. For the dynamic EMI source (SM+BB), we show the EDITER correction with dynamic PE grouping and without grouping - where a single impulse response is calculated for all PE lines.*

***<u>Uncontrolled EMI settings</u>***

*In vivo* images were acquired in 3 subjects with 47.5mT Vanderbilt scanner in the "open" and "flexible shielding" configuration (Figure 4C-D). The scanner room itself was unshielded and contained several uncontrolled EMI sources; usage of devices such as laptops and mobile phones were not restricted within the room to provide a realistic setting. Figure 7 shows PD-weighted images of Subject 1 (male, 25 y/o) in the "open" configuration. Figure 8 shows images of the same subject but in the "flexible shielding" configuration. Figures 7 and 8 show the EMI corrected images using each EMI detector individually (the pickup coil and electrode) and combined. Comparing the detector performance with metric 2 (eq. 4), the pickup coil removed 37.2% and 23.2% of EMI, the electrode removed 89.9% and 64.6% of EMI, and the combination removed 90.2% and 99.9% of EMI in the "open" and "flexible shielding" configuration respectively. Results in 2 other subjects are shown in Supporting Information Figures S3 and S4 to demonstrate repeatability. Table 1 details the standard deviation for corrected and uncorrected data for both the flexible and open shielding configurations for the three subjects.

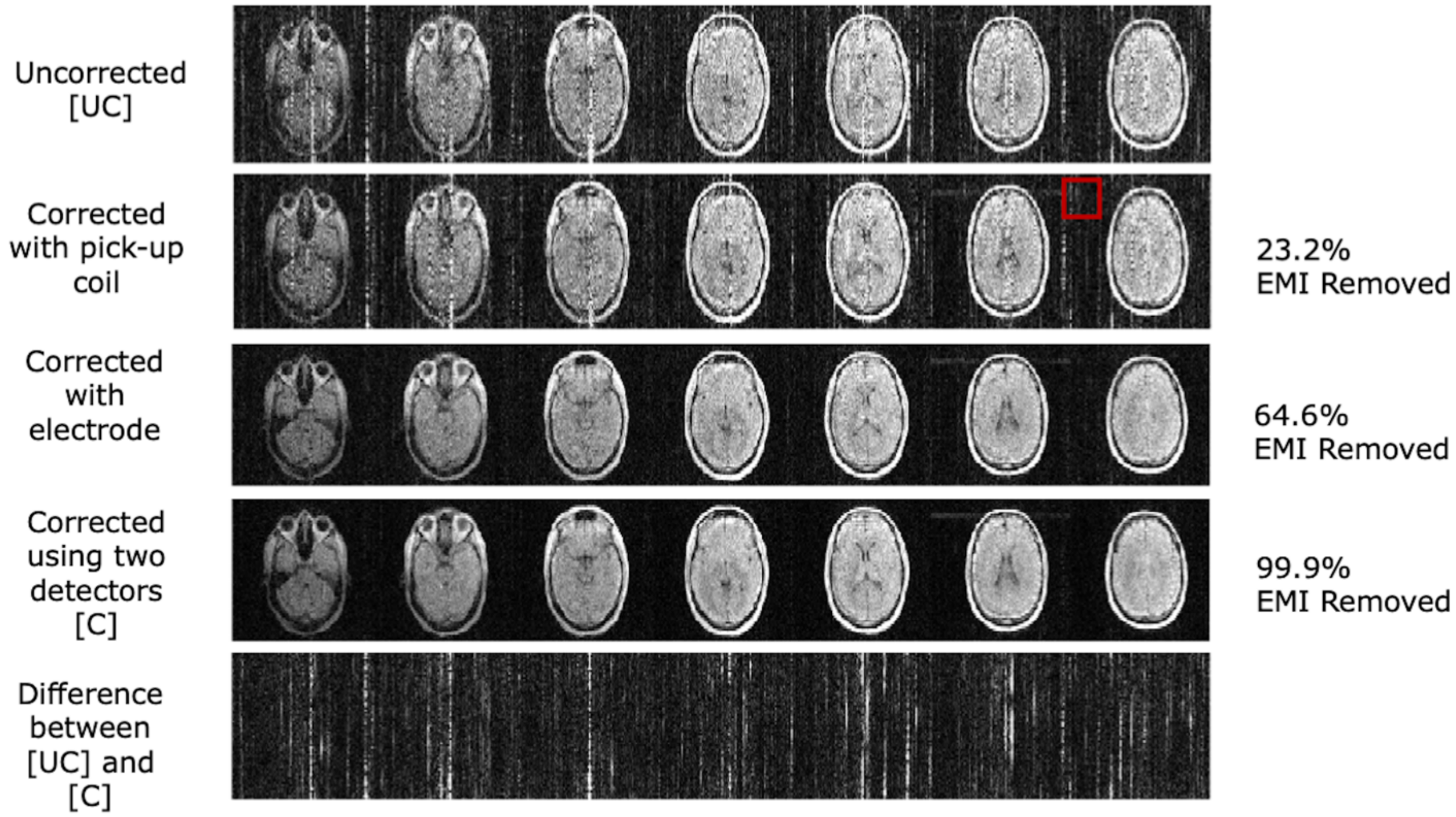

***Figure 7:*** *PD-weighted images using a 3D RARE sequence in the 47.5 mT scanner flexible shielding configuration setup (with draped copper mesh) for Subject 1 (male, 25 y/o). EDITER correction is shown using each EMI detector individually and together. Percent EMI removed by the EDITER method (metric 2) is indicated on the right. The red box indicates the region outside the object that was used for this measurement.*

In the presented results, we used a EMI convolution window size of $\Delta k_x = 7$, $\Delta k_y = 1$. The $N_G$ used per slice for all results shown for three subjects are provided in the Figures. The choice of convolution window widths, $\Delta k_x$, is analyzed in Figure 9. Figure 9A shows the EMI removal percentage (eq. 4) versus $\Delta k_x$. Figure 9B shows the computation time averaged across all subjects versus $\Delta k_x$. Results show that increasing the convolution window width improves the EMI correction performance but increases computation time. We see diminishing returns on the EMI correction for $\Delta k_x$> 4 and a steep increase in computation time for $\Delta k_x$> 8, suggesting $\Delta k_x$<8 as a good choice for the algorithm. In this analysis and in the imaging results, we set $\Delta k_y = 1$. While a 2D convolution window is possible, but we did not observe any advantage in performance with $\Delta k_y$> 1.

## Discussion

The proposed dynamic EMI mitigation method is efficient and straightforward to integrate with POC imaging systems. It was successfully applied to remove EMI corruption in phantom imaging and *in vivo* imaging in two different low-field MRI systems. In a comparison to "ground truth" phantom images our method achieved a total image space RMSE reduction of up to 95.7% (metric 1) and up to 97.3% reduction in the standard deviation of the image background (metric 2).

We demonstrated up to a 99.9% decrease in EMI corruption in *in vivo* imaging results (metric 2). The three *in vivo* imaging datasets were acquired over the course of 57 days and are corrupted by different environmental EMI levels and patterns. Table 1 shows the lowest EMI was achieved with the flexible shielding and EDITER correction. On average, compared to the open shielded uncorrected data, the EDITER correction alone decreased EMI by 71.8%, the flexible shielding alone decreased EMI by 67.3%, and combined they decreased EMI by 89.9%. The flexible shielding alone was much more effective in some cases than others. For example, Subject 2 uncorrected EMI had a relative standard deviation (RSD) of 44.9% in the open configuration and a RSD of 10.7% in the flexible shielding configuration. On the other hand, even with the flexible shielding, Subject 3's uncorrected EMI level had an RSD of 33.2%, but this reduced by 78.8% to RSD = 7.04% with EDITER method.  This disparity could be related to the characteristics of the EMI or from variability in setting up of the flexible shielding mesh.

The dynamic method could be useful in mitigating spatio-temporally varying EMI in new portable MRI settings such as in ambulances or at the bedside. This is accomplished by grouping phase encode lines into correlated temporal windows and creating separate impulse responses ($\vec{h}$) for each group. Instead of PE grouping, we could estimate separate impulses responses for each PE line. However, the use of several PE lines corresponds to longer EMI convolution matrix ($E$) and a better conditioned impulse response estimation which is more robust to noise and provides a better correction overall. In fact, for static/stationary EMI, the correction is done by grouping all PE lines together for calculation of a single $\vec{h}$ or the dataset. The effectiveness of using the dynamic PE grouping as opposed to the static method is illustrated in Figure 6 for the time-varying EMI example. We see an improvement of 75% when using the dynamic PE grouping. We also found that increasing the convolution window size to $\Delta k_x$> 1 aids the correction up to a certain point but can require longer processing time and thus there is a tradeoff between amount of correction and time utilized (figure 9). Ideally a convolution window size 3<$\Delta k_x$ <8 should produce results with a realistic time for processing while improving image quality. Although the scanner we validated in vivo results on is stationery, the $N_G$ grouping we see for some of the results is greater than 1. We might expect higher $N_G$ values for truly portable MR scanners.

Detector positioning is quite important to make the EMI mitigation effective. Supporting Information figure S1 shows the EMI spectra as seen by the primary MR coil and the different EMI detectors for the controlled EMI sources in the shielded room. Observing the EMI detector placement in Figure 2D and 2E, EMI detectors 4 and 5 are closest to the single band (frequency generator) EMI source. Not surprisingly, in Figure S1 A we see that those detectors (4 and 5) pick up the highest amplitude signal from that source. Similarly, the stepper motor EMI source is closest to EMI detectors 1 and 2 and the broadband source is closest to detector 3 which fit with the spectra seen in Figure S1 B and S1 C respectively. Finally, the time-varying source is a combination of stepper motor and broadband source, EMI detectors 1, 2 and 3 are closest to these sources and this agrees with the spectra seen in Figure S1 D. In Supporting Information Figure 2, we see the correction resulting from each detector coil individually. We see that different detectors perform better for different sources based on their position with respect to the source. In a real-world scenario, since the EMI source locations and orientations is unknown, this suggests that it will be beneficial to use multiple detectors distributed around the scanner.

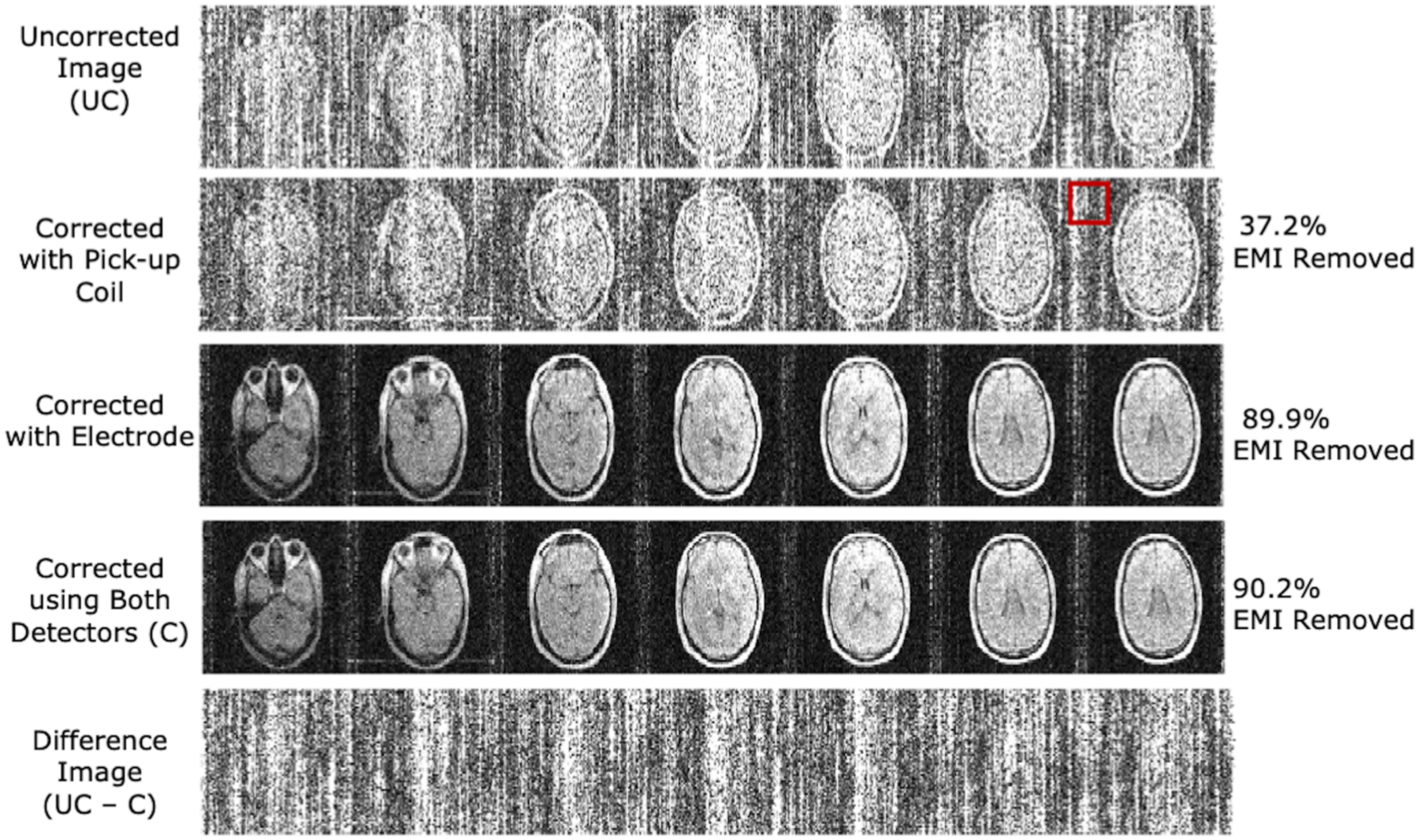

***Figure 8:*** *PD weighted images using a 3D RARE sequence in the 47.5mT scanner's open shielding configuration setup for Subject 1 (male, 25 y/o). EDITER correction is shown using each EMI detector individually and together. Percent EMI removed by the EDITER method (metric 2) is indicated on the right. The red box indicates the region outside the object that was used for this measurement.*

In *in vivo* studies, two different types of external EMI detectors were used, i.e., a tuned and matched coil and an untuned electrode. We predicted that the coil detector would pick up similar EMI to the primary MR coil because of the similar frequency response and detection mechanism. However, as stated, we found that sampled EMI patterns rely heavily on the detector orientation and position with respect to the EMI source. Additionally, the geometry and size of the primary RF coil is different than that of the detector coils which may also cause disparities in the received signal. The electrode on the other hand is non-resonant and picks up EMI induced potentials on the human body which will be presumably "piped in" into the imaging coil regardless of the detectors "positioning" with respect to the EMI sources. Because the electrode is untuned, it likely pick-ups EMI that is also outside the MR acquisition bandwidth and the amplifier system could become saturated by this out-of-band signal. Thus, the use of a band pass filter maybe desirable. We generally found the electrode detector was more effective than the coil detector, but the combined use of both yielded the best results.

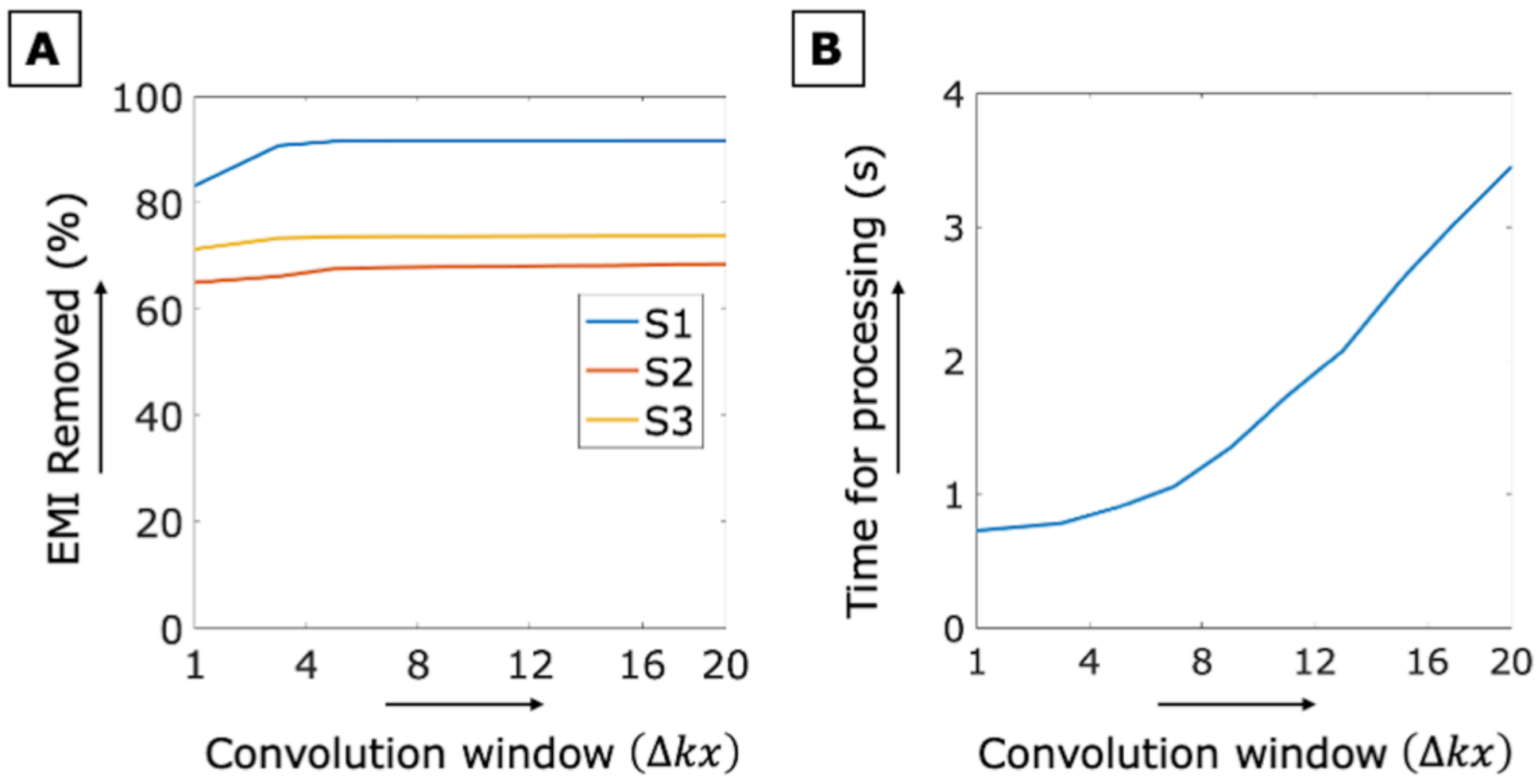

***Figure 9:*** *A) Percent EMI removed (metric 2) versus the convolution window size* $\Delta k_x$ *is shown for the three subjects in the 47.5mT scanner with uncontrolled EMI sources. Results are shown using both EMI detectors in the "open shielding" configuration. B) The average EDITER algorithm processing time versus convolution window size* $\Delta k_x$ *is shown for the "open shielding" configuration scans.*

We use a total of 6 Rx channels for the controlled EMI data and 3 Rx channels for the uncontrolled EMI data which includes 1 Rx channel for the primary MR coil. Using additional Rx channels adds cost and is at odds with the goal of trying to keep spectrometer costs low. Some less expensive options for this include using phase shifters or time domain multiplexing [37] for the received signal in order to receive data from one signal Rx channel. Additionally, a low-cost external receiver with a low sampling rate can be used wherein the signal from the Rx channels can be mixed down and read into the spectrometer using a single Rx channel.

## Conclusion

In this work an efficient retrospective method for EMI mitigation for temporally varying EMI for point-of-care MRI was presented. No additional scan time or pre-scan EMI calibration are required since all the EMI data is acquired simultaneously with the primary MR coil by the EMI detectors. We make use of a shielded room and controllable EMI sources to provide a direct comparison between a "ground truth" image versus our EMI mitigation images with our method. We also demonstrate our method with *in vivo* imaging in 3 subjects. We show that different types of EMI detectors can be used such as RF pick up coils and electrodes and that the latter can directly pick-up EMI from the subject which can account for EMI originating from various directions that may not be effectively picked up by RF coils which are oriented to specific directions. We effectively demonstrated this method on two separate low field scanners to prove that this is a general method that can be retrofitted into scanners. The addition of the EDITER method to compact MRI systems could help bridge the gap to truly portable MRI systems that can be easily sited and re-sited anywhere without a shielded room by mitigating one of the most critical problems in POC low field imaging where SNR is of the essence and passive shield options are limited.

## Acknowledgements

The authors would like to acknowledge Matthieu Sarracanie, Najat Salameh, Cristen LaPierre, Matthew S Rosen and Neha Koonjoo for their past work on the spiral head volume coil and assistance with our implementation of the coil. We thank Monika Sliwiak for help with 3D design and mechanical construction of RF coil assemblies. We acknowledge Patrick C McDaniel for his role in the developing the 80mT portable scanner. The research reported in this publication was supported by the National Institutes of Health grants R01EB018976, 5T32EB1680 and R01 EB030414.

**Code availability.** The MATLAB code for the algorithm is available at https://github.com/abithasrinivas/EDITER_LowfieldMRI.

**Tables:**

| **Subject** | **Flexible Shielding ($\sigma \times 10^6$)** | | **Open Shielding ($\sigma \times 10^6$)** | |
|---|---|---|---|---|
| | Uncorrected | Corrected | Uncorrected | Corrected |
| Subject 1 | 17.5 | 5.36 | 78.9 | 20.2 |
| Subject 2 | 9.92 | 6.09 | 41.5 | 13.9 |
| Subject 3 | 30.7 | 6.49 | 57.1 | 16.2 |

**Table 1:** Standard Deviation of EMI in the ROI outside object (see red box in Fig. 7) averaged across partitions for all subjects in the open and flexible shielding configurations for the uncorrected and corrected images.

**Supporting information Figures**

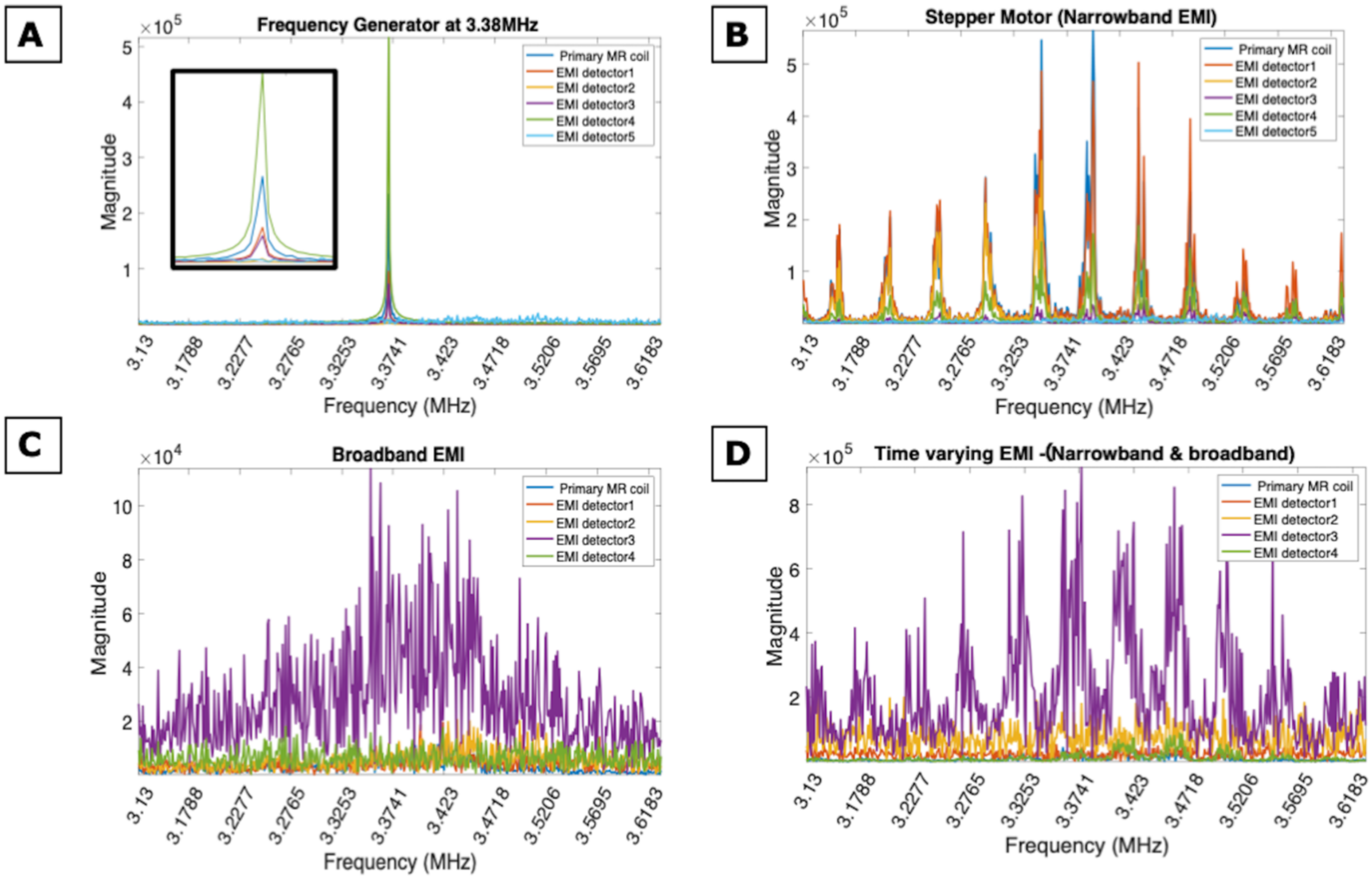

**Supporting information Figure S1:** EMI frequency spectrum received through the primary coil and each EMI detector coils in the portable 80mT system. EMI sources included: A) function generator EMI at 3.38MHz (operating frequency of the portable 80mT low field head scanner) B) Stepper motor (SM) EMI C.) Broadband (BB) EMI source D.) Dynamic switching between the SM and BB sources (narrowband stepper motor source on during the first half of the scan and broadband EMI source on during the second half).

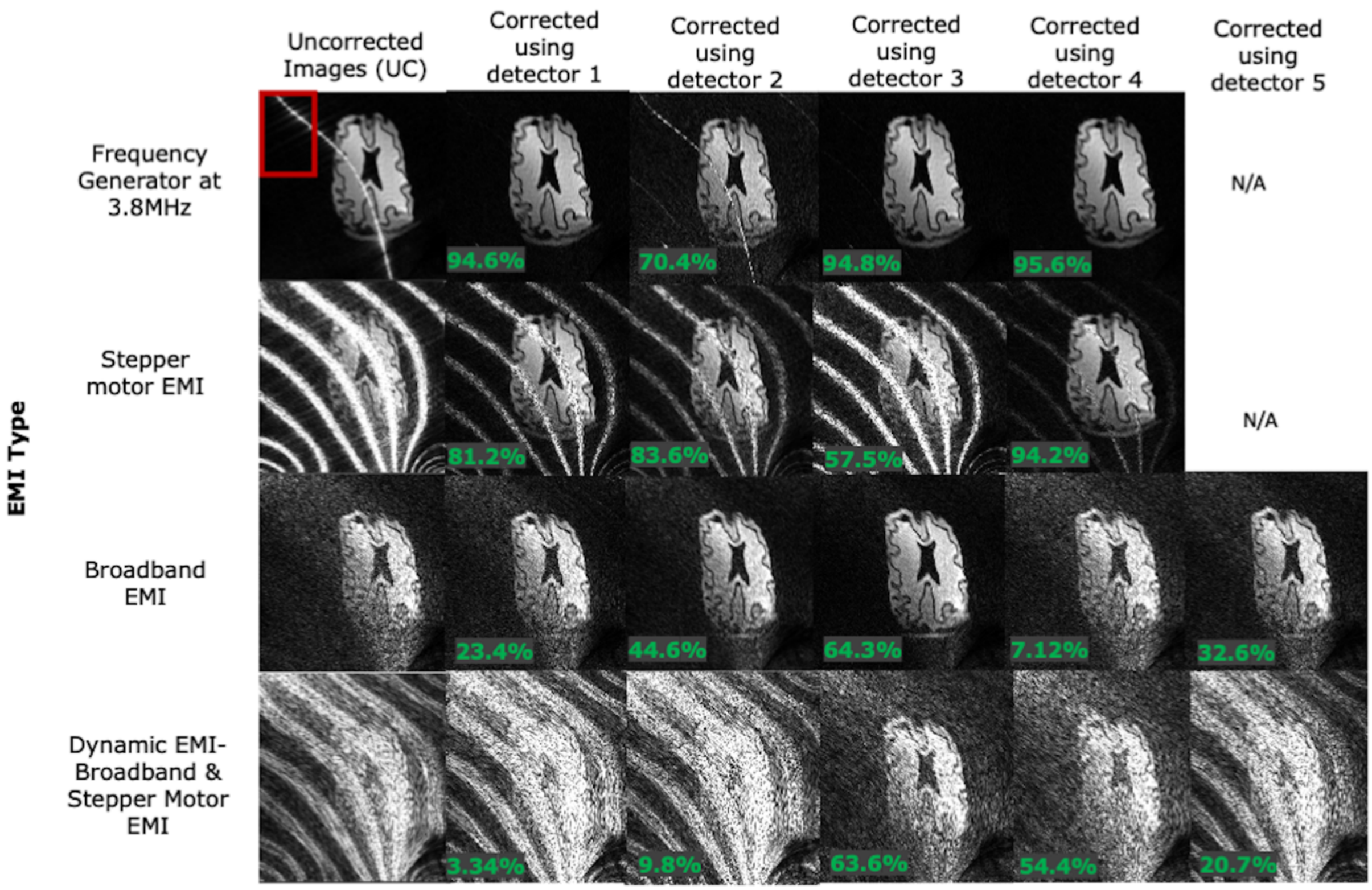

**Supporting information Figure S2:** EMI correction contribution of each detector coil in the portable 80mT scanner is illustrated. 2D images of the brain slice phantom are shown corrected with each detector individually. Percent EMI removed by EDITER (metric 2) is indicated on the bottom left. The red box indicates the region outside the object used to determine this metric.

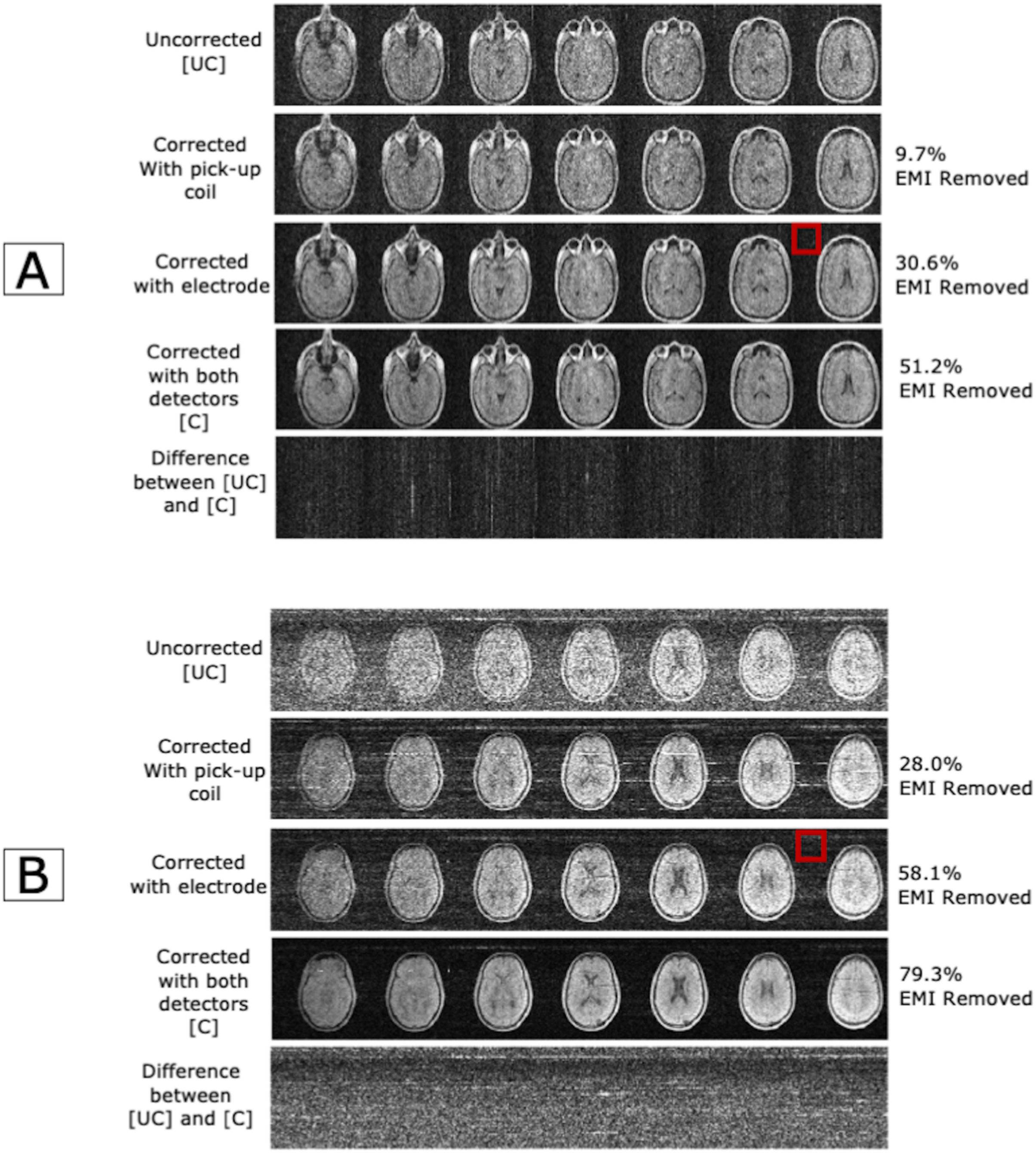

**Supporting information Figure S3:** A) PD weighted images using a 3D RARE sequence in a “flexible” shielding configuration setup (with draped copper mesh) for Subject 2 (female, 27 y/o). B) and for subject 3 (male, 30y/o). EDITER correction is shown using each EMI detector individually and together. Percent EMI removed by the EDITER method (metric 2) is indicated on the right. The red box indicates the region outside the object that was used for this measurement.

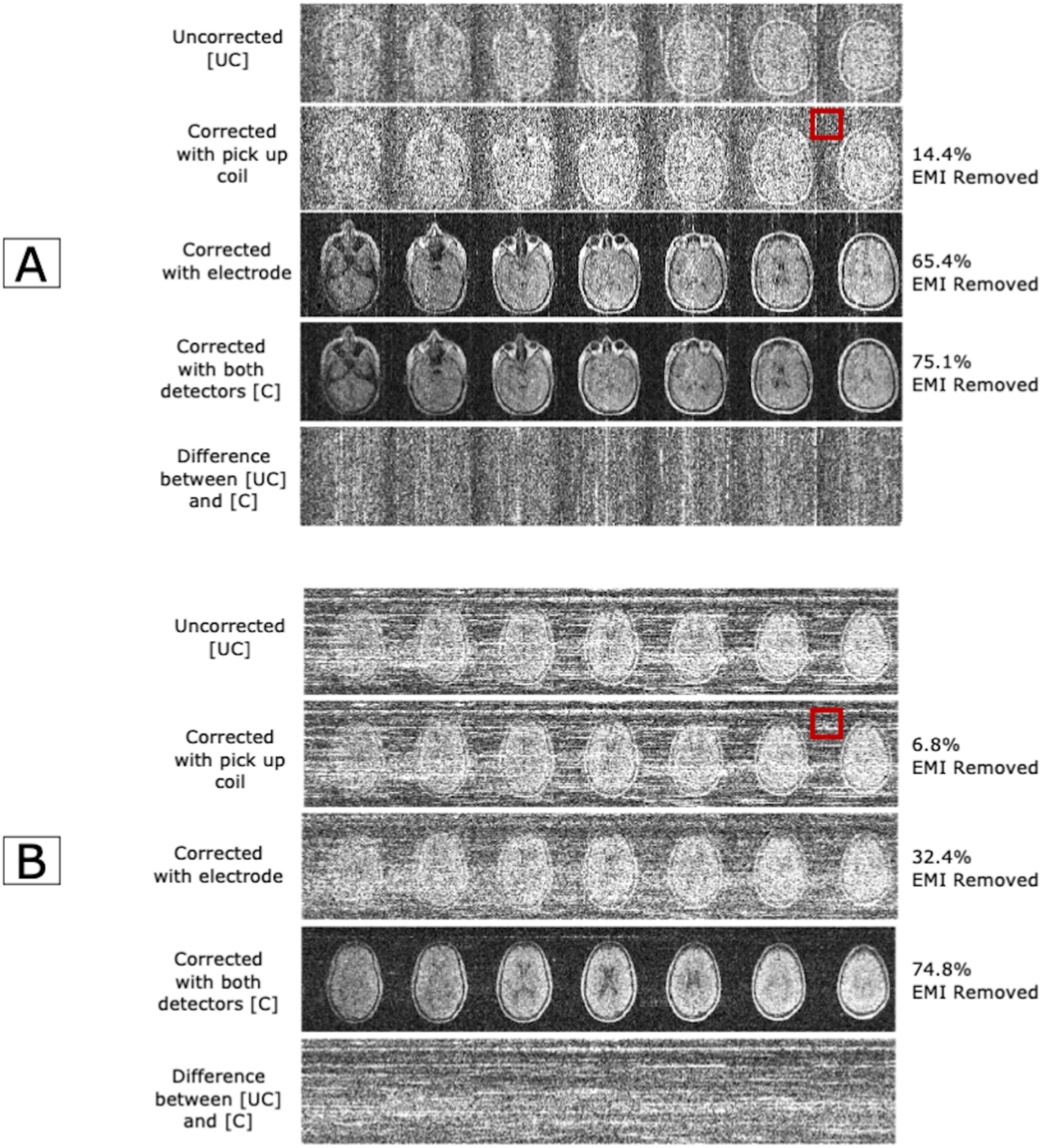

**Supporting information Figure S4:** A) PD weighted images using a 3D RARE sequence in the 47.5mT scanner's "open" shielding configuration setup for Subject 2 (female, 27 y/o). B) and for subject 3 (male, 30y/o). EDITER correction is shown using each EMI detector individually and together. Percent EMI removed by the EDITER method (metric 2) is indicated on the right. The red box indicates the region outside the object that was used for this measurement.